\documentclass[8pt,aps]{revtex4}
\usepackage{amssymb}
\usepackage{latexsym}

\usepackage{epsfig}

\usepackage{dcolumn}
\usepackage{amsmath}
\usepackage{amsfonts}
\usepackage{bm}

\newcommand{\be}{\begin{equation}}
\newcommand{\ee}{\end{equation}}
\newcommand{\bea}{\begin{eqnarray}}
\newcommand{\eea}{\end{eqnarray}}

\newcommand{\p}{\partial}
\newcommand{\s}{\sigma}

\newcommand{\la}{\langle}
\newcommand{\ra}{\rangle}
\newcommand{\rd}{\mbox{d}}
\newcommand{\ri}{\mbox{i}}
\newcommand{\re}{\mbox{e}}

\begin{document}
\title{Universal Features of the Excitation Spectrum in Generalized Gibbs Distribution Ensemble. }

\author{  E. Demler and A.  M. Tsvelik$^*$}
\affiliation
{ Harvard-MIT Center for Ultracold Atoms, Cambridge, MA\\
$^*$Department of  Condensed Matter Physics and Materials Science, Brookhaven National Laboratory, Upton, NY 11973-5000, USA}

\begin{abstract}

It is shown that excitation spectra of Generalized Gibbs Ensembles (GGE) of one-dimensional integrable models with isotopic symmetry contain universal features
insensitive to details of the distribution. Namely, the low energy limit of the subsystem of isotopic (for instance, spin) excitations is described by  the effective action of a ferromagnet at thermodynamic equilibrium with a single temperature and with the stiffness  determined by the initial conditions. The condition of universality is that the entropy per excited particle is small. 

\end{abstract}


\maketitle

\section{General}

 Dynamics of closed interacting many-body systems has lately became a subject of intense interest. Systems can be prepared in out-of-equilibrium state by any non-adiabatic change of the Hamiltonian (for instance, by quench) or by pumping.  In 
such  experiments many-body system is prepared in a state which is not an eigenstate of the Hamiltonian. It is then allowed to evolve coherently with many eigenstates  evolving with different energies. In this paper we will focus exclusively on the fate of integrable systems. 

After some evolution time, which for integrable systems we expect to be of the order of the energy of the lowest energy excitation,
the system can be considered as completely dephased. Then when calculating expectation value of any physical operator,
it is reasonable to neglect off-diagonal terms between different eigenstates since such off-diagonal terms oscillate very
rapidly in time and cancel each other. To be more rigorous one may consider averaging of expectation values
over small time interval. Then  it becomes  possible to describe
the system not as a pure state, but as a density matrix corresponding to some Generalized Gibbs Ensemble (GGE). Physical
properties of the resulting GGE are determined by the integrals of motion which  expectation values in the initial
state fully characterize it \cite{rigol},\cite{cardy},\cite{all},\cite{caux},\cite{essler}. Conditions for validity of GGE for general integrable systems have been recently discussed in \cite{fioretto}.

{\it A priory} there is no reason to expect any universality of resulting GGE. However, in the
next sections we will demonstrate that GGEs appearing in quench experiments with one-dimensional integrable models with
isotopic symmetry, have a universal structure that  are insensitive to details of the distribution. Moreover, the low energy limit
of the subsystem of isotopic (for instance, spin) excitations is corresponds to a ferromagnet at thermodynamic equilibrium
with a single temperature.

 The defining  feature of intergrable systems is existence of infinite set of mutually commuting operators $I_j$ (integrals of motion). Alongside with  the momentum operator $I_0 = P$ and the Hamiltonian $I_1 = H$ these  conserving quantities can be extracted by expanding  the logarithm of transfer matrix in spectral parameter $\theta$:
 \bea
 \ln T(\theta) = P + \ri\theta H + (\ri\theta)^2 I_2/2 +...
 \eea
 Generalized Gibbs Ensemble  (GGE)  described above is given by:
 \bea
 P = Z^{-1}\exp\Big(-\sum_j\beta_j I_j\Big),
 \eea
 where $I_j$ are integrals of motion and $\beta_j$ are corresponding Lagrange multipliers which value is determined by the initial values of $I_j$. It maximizes entropy while taking into account the constraints imposed by the conserved quantities \cite{rigol}, \cite{jaynes}. This hypothesis has been proven in some particular cases \cite{essler},\cite{cassidi}.

 In this paper we consider several typical examples of GGE in relativistic massive field theories possessing some isotopic symmetry. To work with such models is easier because their integrals of motion are known explicitly being fixed by the Lorentz symmetry and requirements of locality. However, as questions may emerge from the point of view of practicality of such models, we will first address prospects of their experimental realization.

  \section{Models}

   In this Section we give several examples of  integrable models which find applications in cold atom physics. Some of them appear in the context of   one-dimensional Bose gas. 
The situation relevant to our discussion is an experiment where a one-dimensional trap  containing  condensate is abruptly  split to become a double-well potential as, for instance, was done in \cite{gring}.  We will be interested in the situation when the split is not complete such that there is still a substantial inter-well tunneling. In the low energy limit interacting 1D Bose condensate is uniquely characterized by its local phase field $\phi(t,x)$ which dynamics is described by the Tomonaga-Luttinger liquid model (the model of non-interacting bosonic field).  When the trap is split there are two condensates and an additional term appears in the Hamiltonian which describes their coupling. As has been demonstrated (see e.g. \cite{Gritsev2007}), the Hamiltonian describing a system of two parallel traps consists of two independent parts. One describes the symmetric part of the condensate phase field $\phi_+ =(\phi_1 + \phi_2)/\sqrt 2$ and the other one describes  the asymmetric part $\phi_- = (\phi_1 - \phi_2)/\sqrt 2$:
\bea
&& H = H_+ + H_- , ~~ H_+ = \frac{v}{2}\int_0^L \rd x \Big[K(\p_x\phi_+)^2 + K^{-1}(\p_x\theta_+)^2\Big]\label{TL}\\
&& H_- =  \frac{v}{2}\int_0^L \rd x \Big[K(\p_x\phi_-)^2 + K^{-1}(\p_x\theta_-)^2 - \lambda\cos(\sqrt{2\pi}\phi_-)\Big]. \label{sG}
 \eea
 where $v$ is the phase velocity, $K$ is the Luttinger parameter normalized in such a way that in Tonks-Girardeau limit $K=1$, and parameter $\lambda$ is proportional to the inter-trap tunneling. Fields $\phi_a, \theta_a$ obey the standard commutation relations:
 \bea
 [\phi_a(x),\theta_b(y)] = \delta_{ab}\Theta_H(x-y),
 \eea
 where $\Theta_H(x)$ is the Heaviside function. The low energy limit of the original bosonic fields is given by
 \bea
 \psi_{1,2} = \sqrt\rho_0 \re^{\ri\sqrt{\pi/2}\phi_+}\re^{\pm \ri\sqrt{\pi/2}\phi_-} + ..., \label{psi}
 \eea
 where dots stand for operators with higher dimensions.

  Models (\ref{TL},\ref{sG}) are integrable; the first one describes a free field, the second one is the famous sine-Gordon model. For $K >1/4$ the cosine term in (\ref{sG}) is relevant and scales to strong coupling. As a result the  sine-Gordon model has massive spectrum of the relativistic form:
  \bea
  E_n(p) = \sqrt{(vp)^2 + M_n^2}.\label{spectrum}
  \eea
Its excitations include solitons and antisolitons corresponding to kinks interpolating between neighboring minima of the cosine potential and (for $K > 1/2$) their bound states called breathers. Their masses in term of the soliton mass $M_0$ are
\bea
M_n = 2M_0\sin\Big[\frac{\pi n}{2(4K-1)}\Big], ~~ n=1,2,.. [4K-1].
\eea
Sine-Gordon model (\ref{sG}) with an additional chemical potential field also describes the superfluid-insulator transition\cite{Buchler2003}. This fact is important for practical applications since in cold atom physics there are well developed techniques for observation of such transition and of its associated features\cite{Bloch2008}.

Another relevant model of ultracold atom physics is the model of SU(N)
fermions with a point-like attractive interaction.  The corresponding Hamiltonian is
\bea
H -\mu N = \int_0^L\rd x\Big( \frac{1}{2m}\p_x\psi^+_a\p_x\psi_a - \mu\psi_a^+\psi_a - g\sum_{a\neq b}\psi^+_a\psi_a\psi^+_b\psi_b\Big) \label{ferm}
\eea
where $a,b = 1,2,...N$. Experimentally one dimensional
systems of SU(2) fermions have been realized by Moritz {\it et al.} \cite{Moritz2005}.
Fermions with SU(N) symmetry have been recently realized in optical lattices in
\cite{Campbell2009,Taie2010}. Earlier theoretical work on SU(N) fermions focused
on their equilibrium properties \cite{Wu2003,Hermele2009,Honerkamp2004,Gorshkov2010}
and their applications to quantum information processing\cite{Gorshkov2009}.

In the limit of weak coupling  $g << (\mu /m)^{1/2}$ model (\ref{ferm}) becomes equivalent to another famous integrable model, the the so-called Chiral Gross-Neveu one. In this limit one  can linearize the spectrum of fermions near the Fermi points and replace the operators:
\bea
\psi(x) = \re^{-\ri k_F x}R(x) + \re^{\ri k_F x}L(x), \label{decomp}
\eea
where slow fields $R,L$ contains Fourier harmonics with momenta much smaller than the Fermi momentum $k_F$. Substituting (\ref{decomp}) into (\ref{ferm}) one obtains 
 \bea
 H = \int_0^L \rd x\Big(-\ri v R^+_{\alpha}\p_xR_{\alpha} + \ri v L^+_{\alpha}\p_x L_{\alpha} - gR^+_{\alpha}L_{\alpha}L^+_{\beta}R_{\beta}\Big),\label{GN}
 \eea
 where $v = k_F/m$.

 Both models (\ref{ferm}) and its relativistic limit (\ref{GN}) are integrable and together with the sine-Gordon model (\ref{sG}) are among the best studied models of that kind. At $g>0$ model (\ref{GN}) is asymptotically free with the interaction scaling to strong coupling in the infrared limit. The spectrum of (\ref{GN}) is split into two independent parts, one of which remains gapless and will not be discussed. The spin sector having SU(N) symmetry have massive excitations with spectrum (\ref{spectrum}) consisting of the fundamental particle with mass  $M_0 \sim \mu (g/v)^{1/N}\exp(- 2\pi v/Ng)$ and its bound states
 \bea
 M_n = M_0\frac{\sin(\pi n/N)}{\sin(\pi/N)}, ~~ n=1,2,...N_1.
 \eea
 These excitations all carry isotopic indices of the SU(N) group and transform according to its fundamental representations described by vertical column of Yang tableau of length $n$.

 Model (\ref{GN}) is particularly interesting for us since it has a non-Abelian continuous symmetry which plays a principal role in  the subsequent discussion.

 \section{TBA equations for GGE}

  Models (\ref{sG},\ref{GN}) were among the first field theories solved by the Bethe Ansatz (BA). To simplify our  consideration we  discuss in detail only the simplest case of SU(2) invariant Gross-Neveu (GN) model $N=2$ and also touch on the sine-Gordon model (\ref{sG}). 

\subsection{General facts about Bethe ansatz}

 Relativistic (Lorentz invariant) models are particularly convenient for our discussion since they admit a simple classification of integrals of motion. This classification becomes particularly transparent in the so-called rapidity representation when energy and momentum of particle of mass $M$ are parametrized as 
\be
E= M\cosh\theta, ~~ P = M\sinh\theta, ~~ E^2 - P^2 = M^2,
\ee
(we set $v=1$ for simplicity). Then Lorentz transformation (boost) becomes just a shift of rapidities of all particles: $\theta_i \rightarrow \theta_i + \alpha$. Consequently, all integrals of motion can be classified according to their Lorentz spin:
\bea
 E \pm P= M\sum_{i =1}^{\cal N}\re^{\pm \theta_i}, ~~ I_{l+1}^{(0)} \pm I_{l+1}^{(1)}= M\sum_{i=1}^{\cal N}\re^{\pm (l+1)\theta_i}, ~~l=1,2,...\label{integrals}
\eea
 
Being a Lorentz invariant object the $N$-body scattering matrix depends only on difference of rapidities of individual particles. For integrable models such $S$-matrix can be written as a product of two-body scattering matrices. Until the system is somehow restricted (for instance, placed in a box), particles rapidities in (\ref{integrals}) are arbitrary. However, as soon as motion of the particles is restricted, this changes. 

  The Bethe ansatz (BA) equations which determine eigenvalues of all integrals of motion for a model in a box of length $L$ with periodic boundary conditions can be derived from the straightforward solution \cite{Andrei} or taking the SU(N)-invariant solution of the Yang-Baxter equations for the two-particle scattering matrix and applying the methods of factorized scattering \cite{ZamZam}. The condition for the periodicity of the wave function of relativistic interacting particles of mass M is 
\bea
\exp(\ri ML\sinh\theta_i)\vec\xi = \prod_{j\neq i}^{\cal N} \hat S(\theta_i-\theta_j)\vec\xi, \label{BA}
\eea
where $\vec\xi$ is a vector depending of spin indices of the particles. The meaning of this equation is straightforward: the $i$-th particle going around the system scatters on all others (it does it one-by-one which is the condition of integrability) acquiring a phase factor given by the product of all $S$-matrices on the right hand side of (\ref{BA}). This phase factor is compensated by $\exp(\ri p_i L)$ in the left hand side of this equation.  

 For models with internal (isotopic) symmetry $\hat S$ is a tensor and diagonalization of (\ref{BA}) requires some effort. The result of this diagonalization is the so-called nested Bethe ansatz. For the GN and the sine-Gordon models the result is 
  \bea
  && \exp(\ri ML\sinh\theta_i) = \prod_{j\neq i}^{\cal N} S_0(\theta_i-\theta_j)\prod_{a=1}^{\cal M} e_1(\theta_i - \lambda_a)\label{theta}\\
  && \prod_{i=1}^{\cal N} e_1(\lambda_a- \theta_i) = \prod_{b=1}^{\cal M} e_2(\lambda_a - \lambda_b) \label{lambda}
  \eea
  where for the GN model
  \[
  e_n(x) = \frac{x - \ri n\pi/2}{x+ \ri n\pi/2}
  \]
  and for the sine-Gordon it is
  \[
  e_n(x) = \frac{\sinh\Big[\frac{\gamma}{2}(x - \ri n\pi/2)\Big]}{\sinh\Big[\frac{\gamma}{2}(x + \ri n\pi/2)\Big]}, ~~ \gamma = (4K-1)^{-1}.
  \]
 where  $S_0(\theta)$ is some known function which exact form is not important for the present discussion. Number ${\cal N}$ stands for the number of particles and ${\cal M}$ is related to the spin projection: $S^z = {\cal N}/2 -{\cal M}$. The qualitative difference between the GN and the sine-Gordon model is that in the latter case for $K >1/2$ function $S_0(\theta)$ have poles on the physical strip and there are bound states (breathers). These breathers carry no spin. We have to add that the integrals (\ref{integrals}) with integer $l$ are local (they have integer Lorenz spin $l$); in principle, one can imagine that quench generates non-local integrals with non-integer $l$. However, the further discussion does not depend on whether it is true or not. 

  Generalization of the nested BA equations (\ref{theta},\ref{lambda}) for models with other simple Lie group symmetry follows the standard scheme described, for instance, in \cite{PCF}. For a given simple Lie group one has to modify $S_0(\theta)$ and replace (\ref{lambda}) with a hierarchy of coupled algebraic equations for rapidities $\lambda^{(j)}$ (j =1,...N-1) where $N-1$ is the dimension of the corresponding Kartan subalgebra. The structure of the  hierarchy reflects the structure of the Dynkin diagram for the given group.

\subsection{Derivation of the universal dynamics}

     The emergence of the universal spin dynamics can be ascertained already from (\ref{theta},\ref{lambda}).  Since the particles are massive, their number in the ground state is zero: ${\cal N} = {\cal M}=0$. However, a nontrivial  GGE emerges after a work had been performed on the system resulting in  a finite particle density. For the infinite system the values of all integrals of motion are determined by the rapidities $\theta_i$; the auxiliary variables $\lambda$ appear only when the system is put in a box. Therefore, at least in the limit of small particle density, one can neglect a feedback on $\theta$'s from $\lambda$'s and consider the distribution of  $\theta$'s as an independent function.

When the ratios $I_j/L$ are finite in thermodynamic limit the distribution function of rapidities $\theta$ must decay sufficiently fast at infinity. Then   in the limit of  large $\lambda$'s one can replace in Eq.(\ref{lambda}) 
\be
\prod_{i=1}^{\cal N} e_1(\lambda_a- \theta_i) \approx [e_1(\lambda_a)]^{\cal N} \label{subs}
\ee
which indicates  that the spin sector (described by $\lambda's$) decouples from $\theta$'s. In fact Eq.(\ref{lambda}) with substitution (\ref{subs}) resembles the BA equation for a spin S=1/2 Heisenberg magnet. In order to figure out whether this is ferro- or antiferromagnet more detailed analysis is needed (see below). As we will show, it is ferromagnet. The difference between the SU(2) GN model and the sine-Gordon model is evident already at this stage: in the former case we have an isotropic magnet and in the latter case it is anisotropic (U(1) or easy plane magnet). 

    In order to see the emergence of universal spin dynamics and establish its conditions we have to derive Thermodynamic Bethe Ansatz (TBA) equations. This derivation follows the standard scheme. First, we establish that generically complex solutions of Eqs.(\ref{lambda}) in the thermodynamic limit ($L \rightarrow \infty, {\cal N}/L, {\cal M}/L = finite$) have only fixed imaginary parts. More specifically, these solutions group into clusters with a common real part (the so-called "strings"):
    \be
    \lambda_{n,j;\alpha} = \theta_{\alpha}^{(n)} + \ri\pi(n + 1 -2j)/2 + O(\exp(-\mbox{const} L)), ~~n=1,2,...; ~ j = 1,2,...n.
    \ee
    Then  we introduce distribution functions of rapidities of string centers $\rho_n(\theta)$ and the distribution function of particle rapidities  $\rho_0(\theta)$. Functions $\tilde\rho_n(\theta), \tilde\rho_0(\theta)$ describe distribution of unoccupied spaces. Their ratios are parametrized by excitation energy functions $\epsilon_n$:
 \be
 \tilde\rho_n(\theta)/\rho_n(\theta) = \exp[-\epsilon_n(\theta)], ~~ \tilde\rho_0(\theta)/\rho_0(\theta) = \exp[-\epsilon_0(\theta)],.
 \ee
\be
{\cal N}/L = \int \rd\theta\rho_0(\theta), ~~ {\cal M}/L = \sum_{n=1}^{\infty}n\int \rd\theta \rho_n(\theta).
\ee
 The entropy of the state is given by the expression
 \be
 S = L\sum_{n=0}^{\infty}\int \rd\theta \left[ (\rho_n + \tilde\rho_n)\ln(\rho_n + \tilde\rho_n) - \rho_n\ln\rho_n - \tilde\rho_n\ln\tilde\rho_n\right].
 \ee
 The TBA equations are a result of minimization of the generalized free energy
 \bea
 \Omega = \sum_j\beta_jI_j - S  = \int \rd\theta K(\theta)\rho_0(\theta)  - S
 \eea
where 
\bea
K(\theta) = \sum_{k}[\beta_k\re^{k\theta}+\bar\beta_k\re^{-k\theta}], \label{tba2}
\eea 
  subject to constraints imposed by the equations for the distribution functions:
 \bea
  \tilde\rho_n + \rho_n = s*(\tilde\rho_{n-1} + \tilde\rho_{n+1}) + \delta_{n,0} \frac{M}{2\pi}\cosh\theta, \label{tba3}
  \eea
  The result is
  \bea
  && \epsilon_n(\theta) = s*\ln[1+ \re^{\epsilon_{n-1}(\theta)}][1+ \re^{\epsilon_{n+1}(\theta)}] - \delta_{n,0}K(\theta) \label{tba1}\\
  &&\Omega/L = - \int\frac{\rd\theta}{2\pi}K(\theta)\ln\Big[1 + \re^{\epsilon_0(\theta)}\Big]\label{F}\\
  && s*f(\theta) = \int_{-\infty}^{\infty}\frac{\rd\theta' f(\theta')}{2\pi \cosh(\theta - \theta')}\nonumber
\eea
From these equations one can restore the integrals of motion:
\bea
I_j = - \frac{\p\Omega}{\p \beta_j}. \label{I}
\eea

 A peculiar property of TBA equations (\ref{tba3},\ref{tba1}) is their quasi-locality:  given $\epsilon_n,\rho_n$ are related only to their nearest neighbors. Therefore, if $\epsilon_0,\rho_0$ are fixed the TBA for $\epsilon_n, n=1,2, ...$ will have the same form as (\ref{tba1}), but with $K$ replaced by $G(\theta) = s*\ln(1+ \re^{\epsilon_0(\theta)})$.

\begin{figure}[ht]
\begin{center}
\epsfxsize=0.5\textwidth
\epsfbox{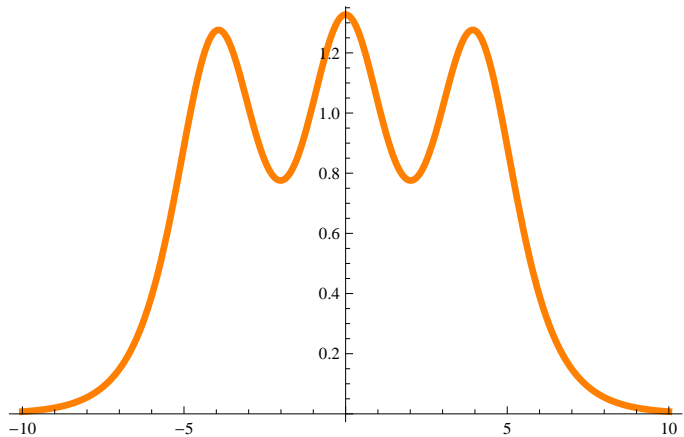}
\end{center}
\caption{An example of integrable function $G(\theta;\{\beta\})$. Function $\epsilon_0$ changes sign six times going to $-\infty$  at both infinities. }
\end{figure} 

\begin{figure}[ht]
\begin{center}
\epsfxsize=0.5\textwidth
\epsfbox{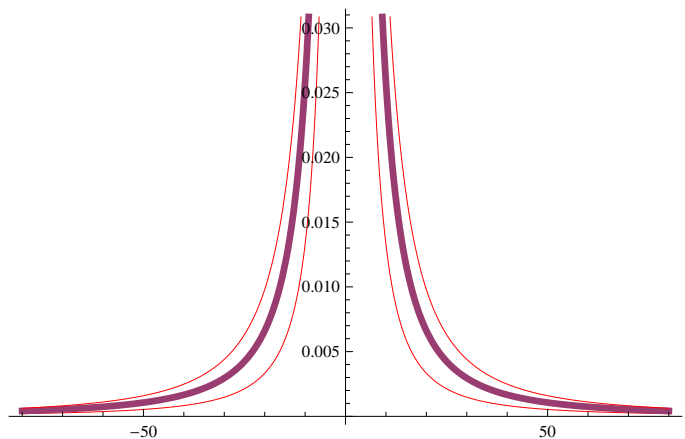}
\end{center}
\caption{$\epsilon_{n}(\theta)$ functions for $n=1,2,3$ corresponding to $G(\theta)$ shown at Fig.1. The asymptotics does not depend on fine details of behavior $G(\theta)$ at finite $\theta$. }
\end{figure} 

   It is natural to assume that all integrals of motion are extensive quantities $\sim L$ and hence their densities are finite. It follows then from (\ref{F}) that  $\epsilon_0(\theta) \rightarrow -\infty$  at infinity  (changing sign more than ones at finite $\theta$). Then  the function $G(\theta)$ is integrable which will be important for what follows. On Fig. 1 we give an example of such function.  We have chosen $\epsilon_0$ in such a way, that $G(\theta)$ has some nontrivial structure at small $\theta$. This structure is determined by the initial conditions of the quench. All these details, however, does not make a difference in in the asymptotic region of large $\theta$ (see Fig. 2) where the bahavior of $\epsilon_n$ is determined by a single integral characteristic of the $\theta$-distribution (see the derivation below). 
 
Inverting the kernels in (\ref{tba3},\ref{tba1}) with $n> 1$ we obtain TBA equations for GGE in the following form:
   \bea
   && \ln[1+ \re^{\epsilon_{n}(\theta)}] - A_{nm}*\ln[1+\re^{-\epsilon_{m}(\theta)}] = a_n*G(\theta), ~~n,m =1,2,...\label{tba5}\\
   && \tilde\rho_{n} + A_{nm}*\rho_{m} = a_n*s*\tilde\rho_0 \label{tba6}
   \eea
   where the Fourier images of the kernels are
   \[
   a_n(\omega) = \exp[- \pi n|\omega|/2], ~~ A_{nm}(\omega) = \coth(\pi\omega/2)\Big[\exp(- |n-m|\pi |\omega|/2) - \exp(-(n+m)\pi |\omega|/2)\Big].
   \]
  Eqs.(\ref{tba5},\ref{tba6}) with $G$ and $s*\tilde\rho_0$ replaced by the delta functions are precisely TBA's for a spin S=1/2 Heisenberg ferromagnet in thermodynamic equilibrium (!) \cite{Ferro}. 
 When these functions are not delta functions, but just sharp peaks, the analogy with the ferromagnet remains valid just asymptotically. For this analogu to hold  it is sufficient that the integral $\int G(y) \rd y$ converges, but is $>> 1$. This would correspond to low effective temperature limit of the ferromangnet when the free energy is determined by large rapidities (small momenta) such that details of the dispersion at large momenta are not important.    Indeed we have
\bea
&& a_n*G(\theta) = n\pi\int\rd y\frac{G(y)}{(\theta-y)^2 + \pi^2n^2} \rightarrow \frac{n\pi}{2 \theta^2}\int \rd y G(y) \nonumber\\
&& J/T = \frac{1}{2}\int \rd y G(y) \label{J}.
\eea
A similar condition must be satisfied by the function $\rho_0(\theta)$:
\bea
\int \rd y\rho_0(y) = \mbox{finite}.
\eea

The condition $\int G(y) \rd y >> 1$ means that there is little feedback from $\epsilon_n, ~~n \geq 1$ to $\epsilon_0$ and therefore one may use function $G(\theta)$ to characterize the state. This is more convenient than to use the intergrals of motion which themselves can be restored from $G$ via (\ref{I}). From (\ref{F}) and (\ref{J}) it follows that the effective exchange integral of our ferromagnet is of the order of the energy density per particle. Since the entropy per particle in the ferromagnet $~ (T/J)^{1/2}$, the requirement $T/J << 1$ is equivalent to  requiring   the entropy per particle to be  small. Under that condition the  universal spin dynamics emerges in GGE which is, perhaps, the most striking result of our derivation. It is also clear that  the spin subsystem of GGE is at thermal equilibrium  and is described by the ordinary Gibbs ensemble with a single temperature.


\section{Correlation functions}

To understand how  the universal dynamics described in the previous subsection is related to observable quantities one has to consider correlation functions.  This is a difficult problem and though we are not in a position to offer a detailed solution, we feel obliged to make some remarks.

To be closer to real experimental systems we consider correlators of  the bosonic creation and annihilation operators in the model of two coupled Bose condensates (\ref{TL},\ref{sG}). In the low energy limit these operators are expressed in terms of the phase fields $\phi_{\pm}$ (\ref{psi}). Since the two sectors of the model are decoupled, the correlation functions factorize. For instance, for the 2-point one we have:
\bea
\la\la \psi_a(t,x)\psi_b^+(0,0)\ra\ra \sim \rho_0\la\la \re^{\ri \sqrt{\pi/2}\phi_+(t,x)}\re^{-\ri\sqrt{\pi/2}\phi_+(0,0)}\ra\ra \la\la \re^{\ri \sqrt{\pi/2}\phi_-(t,x)}\re^{(1-2\delta_{ab})\ri\sqrt{\pi/2}\phi_-(0,0)}\ra\ra
\eea
The correlation functions of bosonic exponents of gapless fields in GGE have been calculated \cite{gapless}, but correlators of the sine-Gordon fields $\phi_-$ have never been analyzed in this context. Below we will discuss some general features of these correlators.

We will discuss the limit of zero effective temperature $T_{eff}=0$ when all $\epsilon_n \rightarrow \infty$ in (\ref{tba5}), but their ratios remain constant.  This is the limit when GGE distribution is reduced to a single "vacuum" state. This state is characterized by some distribution of the particle rapidities and is ferromagnetic, that is has maximal possible spin. Using  the Lehmann expansion where  Green's functions are  expanded in matrix elements over excited states $|n\ra$:
\bea
\la {\cal O}(t,x) {\cal O}^+(0,0)\ra = \sum_n|<n|{\cal O}(0,0)|vac>|^2\re^{- \ri t(E_n - E_{vac}) - \ri x(P_n - P_{vac})}
\eea
In the sine-Gordon model excitations are classified as solitons, anti-solitons and their bound states (see Section II). (Anti)solitons can be described as particles with (negative) positive spin projection ($s = \pm 1/2$). The ferromagnetic state in this context corresponds to the state with only one type of particles (for example, solitons). We denote such state as
\[
|\Big(\theta_n,...\theta_1\Big)_S;\Big(0\Big)_{AS}>
\]
According to \cite{Lukyanov}, the following matrix elements do not vanish:
\bea
&& <\Big(\theta_1,...\theta_n\Big)_S;\Big(0\Big)_{AS}|\re^{\ri\beta\phi_-}|\Big(\theta_1',...\theta'_{n+N}\Big)_S;\Big(\bar\theta_1,...\bar\theta_N\Big)_{AS}> = \nonumber\\
&& <vac|\re^{\ri\beta\phi_-}|\Big(\theta_1',...\theta'_{n+N}\Big)_S;\Big(\bar\theta_1...\bar\theta_N, \ri\pi + \theta_1, ...\ri\pi + \theta_n\Big)_{AS}>
\eea
These matrix elements would correspond to excitation of $N$ magnons and therefore the spectral function of  bosons contains multi-magnon processes. Unfortunately, the form is these matrix elements is very cumbersome which makes further calculations difficult. At the moment the moment this is all we can say. 

\section{Other models}

The results obtained for the SU(2)-invariant model (\ref{GN}) can be easily generalized for any simple Lie group. Qualitative differences appear only when the S-matrix for physical particles contains the so-called RSOS (Restricted Solid-on-Solid) component. Such models have an exotic degenerate ground state and excitations with non-Abelian statistics. In view of rarity of such problems we will confine ourselves to a brief discussion.

A typical representative of this  class of models is the SU$_k$(2) Wess-Zumino-Novikov-Witten (WZNW) model perturbed by the current-current interaction term:
\bea
H  = \int _0^L\rd x \Big[ \frac{2\pi v}{k+2}\Big(:J^aJ^a:  + :\bar J^a\bar J^a:\Big)+ g J^a\bar J^a\Big] \label{wznw}
\eea
where the current operators satisfy SU(2) Kac-Moody algebra of level $k$:
\bea
[J^a(x),J^b(y)] =\ri\epsilon^{abc}J^c(x)\delta(x-y) + \frac{k}{2\pi}\delta^{ab}\delta'(x-y)
\eea
which, in fact, coincides with the commutation relations of the fermionic bilinears
\[
J^a = \sum_{j=1}^kR^+_{j\alpha}\s^a_{\alpha\beta}R_{j\beta}, ~~ \bar J^a = \sum_{j=1}^kL^+_{j\alpha}\s^a_{\alpha\beta}L_{j\beta}
\]
Model (\ref{wznw}) describes the SU(2)-invariant sector of the SU(2)$\times$SU(k) model of fermions \cite{tsvelik}. The soliton excitations of this model carry zero modes of parafermions; this is the origin of the non-Abelian statistics  (see, for instance \cite{smirnov},\cite{fendley} for the discussion). This model also emerges as a continuum limit of the lattice model of spin $S = k/2$ integrable magnet with a small Ising-like anisotropy $\Delta$\cite{fatzam}:
\bea
H = \sum_j {\cal P}_k\Big[(S^+_jS^-_{j+1} + h.c.),S^z_jS^z_{j+1};\Delta\Big]
\eea
where 
\[
{\cal P}_k(x,y;\Delta) = \sum_{n+m \leq k} A_{nm}(\Delta)x^ny^m
\]
is some known polynomial. The coupling constant in (\ref{wznw}) $g \sim \sqrt{\Delta -1}$. 

The TBA equations for this model differ from (\ref{tba1},\ref{tba2},\ref{tba3}) only in one respect: the driving term is placed not in the first, but in the $k$-th equation and the free energy is also determined by $\epsilon_k$\cite{tsvelik}. Therefore inverting the kernels in TBA yields not one, but two sets of independent equations:
\bea
&& \ln[1+ \re^{\epsilon_{n+k}(\theta)}] - A_{nm}*\ln[1+\re^{-\epsilon_{m+k}(\theta)}] = a_n*G(\theta), ~~n,m =1,2,...\\
&& \ln[1+ \re^{\epsilon_{n}(\theta)}] - {\cal A}_{nm}*\ln[1+\re^{-\epsilon_{m}(\theta)}] = {\cal B}_n*G(\theta), ~~n,m =1,2,...k-1\label{zk}
\eea
where
\[
{\cal B}_n = \frac{\sinh[\pi(k-n)\omega/2]}{\sinh(\pi k\omega/2)}, ~~ {\cal A}_{nm} = 2\coth(\pi\omega/2)\frac{\sinh[\pi(k - \mbox{max}(n,m))\omega/2]\sinh[\pi\mbox{min}(n,m)\omega/2]}{\sinh(\pi k\omega/2)}
\]
and the corresponding equations for the densities are
\bea
&& \tilde\rho_{n+k} + A_{nm}*\rho_{m+k} = a_n*s*\tilde\rho_k\\
&& \tilde\rho_{n} + {\cal A}_{nm}*\rho_{m+1} = {\cal B}_n*s*\tilde\rho_k \label{zk2}
\eea
The set of equations for $n >k$ describes spin S=1/2 ferromagnet as before. As far as the equations for $n <k$ are concerned, in the low energy limit they describe a conformal theory. From (\ref{zk}) and (\ref{zk2}) it follows that at large $|\theta|$
\bea
\epsilon_n(\theta) \sim \re^{- 2|\theta|/N}, ~~ \tilde\rho_n(\theta) \sim \re^{-2|\theta|/N},
\eea
and, since $p(\theta) = 2\pi\int^{\theta}\rd\theta' \tilde\rho(\theta')$, the spectrum is linear.
The low energy limit of the second one corresponds to the conformal field theory of Z$_k$ parafermions.

\section{Physical consequencies}

 We conclude this paper by reiterating its main result: provided the initial state of our system containing a gas of excited particles is a low entropy state with (small entropy per particle), its spin dynamics is universal. This condition is equivalent to the condition $T << J$ (the effective exchange integral (\ref{J})).  Namely, in the low energy limit GGE spin subsystem looks like a ferromagnet at thermodynamic equilibrium. GGE's of models with non-Abelian statistics also contain a critical sector described by some conformal field theory (for the example given in the main text it was the theory of Z$_k$ parafermions). A qualitative explanation for this result is simple: quench creates particles not fixing the total magnetic moment and the system chooses a state with maximal entropy which is a state with maximal total moment.

  There two natural questions to ask. The first is how to prepare such a state, the second is how this universal dynamics reveals itself in observable quantities. One possible answer to the first question is to use pumping with an
appropriate field generating pairs of excitations in the spin sector. In the case of
ultracold atoms in optical lattices such pumping can be performed with optical lattice modulation. A pulse with duration $t_0$ and frequency
$\omega$ may excite pairs of particles of type $n$ momenta provided that
$\omega > 2 M_n$. Assuming that pumping does not give momentum to the system,
excitations will be generated as pairs of quasiparticles with opposite momenta. Energy conservation then gives that such quasiparticles will be excited with rapidities centered around $M_n \cosh \theta \approx w$ and the width of energy distribution $t_0^{-1}$.

 Now let us discuss the second question. What are the physical consequencies of the effectively "thermal" character of the spin subsystem?
Obviously, one cannot argue that the form of generic isospin correlation functions will be that of a corresponding ferromagnet. This is clear from the analysis of Section IV. Therefore even though we
find the same distribution of eigenstates as in the appropriate ferromagnet, matrix elements of the operators 
may be very different. Is it possible then to perform any physical experiments, which
would demonstrate the corresponding "thermal" character of the isospin sector?

The fist option is to
measure fluctuations of the magnetization. For the Chiral Gross-Neveu model the
former corresponds to selecting any of the SU(N) operators and measuring the value of this operator in a finite segment of the system\cite{Cherng2007,Eckert2007}. Unlike more generic operators, smooth components of the
spin density operators have the same matrix elements for the Chiral Gross-Neveu and ferromagnetic models
(analogously $N_{-} \sim \p_x\phi_-$ operator for the sine-Gordon has the same matrix elements as the $S^z$ operator for the easy-plane ferromagnet).
We emphasize that measurements of the spin operator should be repeated many times so that one could extract not just the average value but all
fluctuations of the operator. Putting it  differently one can say that results of individual measurements should be combined into  distribution function.
We predict that the distribution function of magnetization fluctuations of the Chiral Gross-Neveu model
after a quench should be the same as in a ferromagnet at finite temperature.

The second way of observing thermal character of the isospin sector is to measure its fluctuations of energy.
In a  thermal ensemble energy fluctuations are given by the specific heat $\langle \Delta E^2 \rangle = c_V T$\cite{Pathria1972}.
In equilibrium specific heat is itself a function of temperature. Hence by measuring the average energy and its fluctuations
one can effectively measure the equation of state. Our analysis suggests that the average energy and fluctuations of the energy
in the Chiral Gross-Neveu model following the quench should be given by the equilibrium equation of state of an SU(N) ferromagnet.
Measurements of energy fluctuations in the system (or in
a fragment of the full system) can  be done experimentally (see e.g. experiments \cite{Kinoshita2004}. In these experiments
the average energy of an interacting 1D Bose gas was measured in an array of tubes. Recent experiments allow local resolution
of individual 1d systems\cite{Folman2002,Bakr2009,Sherson2010}, which should make it possible to measure not only the average energy
but also energy fluctuations. Assuming separation of the isospin and density sectors, it should also be possible so separate
the isospin part of the energy from the total energy.


 AMT is  grateful to  Robert Konik, Andrew Green, Misha Zvonarev and Fabian Essler for interesting and inspiring discussions. We also thank Galileo Galilei Institute where this work was finished for its hospitality. 
AMT was  supported  by US DOE under contract number DE-AC02 -98 CH 10886.
ED acknowledges support from   Harvard-MIT CUA, NSF Grant No. DMR-07-05472,
the Army Research Office with funding from the DARPA OLE
program, AFOSR Quantum Simulation MURI, 
the ARO-MURI on Atomtronics.


\begin{thebibliography}{99}

\bibitem{rigol} M. Rigol, V. Dunjko, V. Yurovsky and M. Olshanii, Phys. Rev. Lett. {\bf 98}, 050405 (2007); Nature {\bf 452}, 854 (2008).
\bibitem{cardy} P. Calabrese and J. L. Cardy, J. Stat. Mech: Th. Exp. {\bf 2007}: P06008 (2007). 
\bibitem{all} M. A. Cazalilla, A. Iucci, M. C. Chang, Phys. Rev. E {\bf 85}, 011133 (2012) 
M. Cramer {\it et.al.}, Phys. Rev. Lett {\bf 100}, 030602 (2008); T. Barthel and Schollw\"ock, ibid. {\bf 100}, 100601 (2008); G. Roux, Phys. Rev. A{\bf 79}, 021608 (2009).
\bibitem{caux} J. Mossel and J.-S. Caux, New J. Phys. {\bf 12}, 055028 (2010); arXiv:1203.1305.
\bibitem{essler} P. Calabrese, F. H. L. Essler and M. Fagotti, Phys. Rev. Lett. {\bf 106}, 227203 (2011).
\bibitem{fioretto} D. Fioretto and G. Mussardo, New J. Phys. {\bf 12}, 055015 (2010). 
\bibitem{jaynes} E. T. Jaynes, Phys. Rev. B{\bf 106}, 620; ibid. {\bf 108}, 171 (1957).
\bibitem{cassidi} A. C. Cassidi, C. W. Clark and M. Rigol, Phys. Rev. Lett. {\bf 106}, 140405 (2011).
\bibitem{gring} M. Gring, M. Kuhnert, T. Langen, T. Kitagawa, B. Rauer, M. Schreitl, D. A. Smith, E. Demler, J. Schmiedmayer, arXiv:1112.0013.
\bibitem{Gritsev2007}
V. Gritsev, A. Polkovnikov, and E. Demler.
Phys. Rev. B {\bf 75}, 174511 (2007).


\bibitem{Buchler2003}
H.~P. B\"uchler, G.~Blatter, and W.~Zwerger.
Phys. Rev. Lett. {\bf 90}, 130401 (2003).

\bibitem{Bloch2008}
I. Bloch, J. Dalibard, and W. Zwerger.
Rev. Mod. Phys. {\bf 80}, 885 (2008).

\bibitem{Moritz2005}
H. Moritz, T. St\"oferle, K. G\"unter, M. K\"ohl, and T.
  Esslinger.
 Phys. Rev. Lett. {\bf 94}, 210401 (2005).

\bibitem{Campbell2009}
G. K. Campbell, M. M. Boyd, J.~W. Thomsen, M. J. Martin, S. Blatt, M. D. Swallows, T. L. Nicholson, T. Fortier, C. W. Oates, S. A. Daddams, N. D. Lemke, P. Naidon, P. Julienne, J. Ye, A. D. Ludlow,
Science {\bf 32}, 360 (2009).

\bibitem{Taie2010}
S. Taie, Y. Takasu, S. Sugawa, R. Yamazaki, T. Tsujimoto, R. Murakami, Y. Takahashi,
Phys. Rev. Lett {\bf 105}, 190401 (2010).

\bibitem{Wu2003}
C. Wu, J-P. Hu, and S-C. Zhang, 
Phys. Rev. Lett. {\bf 91}, 186402 (2003). 

\bibitem{Hermele2009}
M. Hermele, V. Gurarie, and A.-M. Rey.
Phys. Rev. Lett. {\bf 103}, 135301 (2009).

\bibitem{Honerkamp2004}
C. Honerkamp and W. Hofstetter,
 Phys. Rev. Lett. {\bf 92}, 170403 (2004).

\bibitem{Gorshkov2010}
A. V. Gorshkov, M. Hermele, V.~Gurarie, C. Xu, P. S.  Julienne, J. Ye, P. Zoller, E. Demler, M. D. Lukin, A. M. Rey,
 Nature Physics {\bf 6}, 289 (2010).

\bibitem{Gorshkov2009}
A.~V. Gorshkov, A. M. Rey, A. J. Daley, M. M. Boyd, J. Ye, P. Zoller, M. D. Lukin,
Phys. Rev. Lett. {\bf 102}, 110503 (2009).

\bibitem{Andrei} N. Andrei, J. H. Lowenstein,  Phys. Lett B{\bf 90}, 106 (1980); ibid., {\bf 91}, 401 (1983).
\bibitem{ZamZam} A. B. Zamolodchikov and Al. B. Zamolodchikov, Ann. Phys. {\bf 120}, 253 (1979).
\bibitem{PCF} E. Ogievetskii, N. Yu. Reshetikhin and P. B. Wiegmann, Nucl. Phys.{\bf 280}, 45 (1987). 
\bibitem{Ferro} M. Takahashi, Prog. Theor. Phys. {\bf 46}, 401 (1971).
\bibitem{gapless} A.Mitra and T. Giamarchi, Phys. Rev. B{\bf 85}, 075117 (2012).
\bibitem{Lukyanov} S. Lukyanov, A. Zamolodchikov, Nucl. Phys. {\bf 607}, 437 (2001).
\bibitem{tsvelik} A. M. Tsvelick, ZhETP {\bf 93}, 1329 (1987).
\bibitem{smirnov} F. A. Smirnov, Int. J. Mod. Phys. A{\bf 9}, 5121 (1994).
\bibitem{fendley} P. Fendley and H. Saleur, Phys. Rev. D{\bf 65}, 025001 (2002).
\bibitem{fatzam} A. B. Zamolodchikov and V. A. Fateev, Sov. J. Nucl. Phys. {\bf 32}, 298 (1980).


\bibitem{Cherng2007}
R.~W Cherng and E. Demler, 
New Journal of Physics {\bf 9}, 7 (2007).

\bibitem{Eckert2007}
K.~Eckert, \L{}. Zawitkowski, A.~Sanpera, M.~Lewenstein, and E.~S. Polzik, 
Phys. Rev. Lett. {\bf 98}, 100404 (2007).
\bibitem{Pathria1972} P. K. Pathria, P. D. Beale, "Statistical mechanics", third edition, Elsevier, 2011
\bibitem{Kinoshita2004}
T. Kinoshita, T. Wenger, and D.~S. Weiss.
\newblock {\em Science}, {\bf 305} 1125 (2004).

\bibitem{Folman2002}
R. Folman, P. Kr�ger, J. Schmiedmayer, J. Denschlag, and C.
  Henkel.
\newblock volume~48 of {\em Advances In Atomic, Molecular, and Optical
  Physics}, pp. 263 -- 356. Academic Press, 2002.

\bibitem{Sherson2010}
J. F. Sherson, C. Weitenberg, M. Endres, M. Cheneau,  I. Bloch, S. Kuhr,
Nature {\bf 467}, 68 (2010).


\bibitem{Bakr2009}
W. S. Bakr, J. I. Gillen, A. Peng, S. Folling,  M. Greiner, 
Nature {\bf 462}, 74 (2009).






\end{thebibliography}
\end{document}